# SBML FOR OPTIMIZING DECISION SUPPORT'S TOOLS

Dalila Hamami[1] and Baghdad Atmani[2]

Computer science laboratory of Oran (LIO)

[1]Department of Computer Science, Mostaganem University, Algeria
dhamami8@gmail.com
[2]Oran university, Algeria
atmani.baghdad@gmail.com

## ABSTRACT

*Many theoretical works and tools on epidemiological field reflect the emphasis on decision-making tools by both public health and the scientific community, which continues to increase.*

*Indeed, in the epidemiological field, modeling tools are proving a very important way in helping to make decision. However, the variety, the large volume of data and the nature of epidemics lead us to seek solutions to alleviate the heavy burden imposed on both experts and developers.*

*In this paper, we present a new approach: the passage of an epidemic model realized in Bio-PEPA to a narrative language using the basics of SBML language. Our goal is to allow on one hand, epidemiologists to verify and validate the model, and the other hand, developers to optimize the model in order to achieve a better model of decision making. We also present some preliminary results and some suggestions to improve the simulated model.*

## 1. INTRODUCTION

In recent years, biotechnology improved the knowledge of epidemiological pathogens, and developed effective ways to fight against these epidemics. Currently, several outbreaks are in vogue, and developing factors helped build huge data banks [1]. Therefore, the amounts of raw data are too large to be analyzed manually by both experts and computer scientists who need to understand the epidemiological domain.

Due to the variety of biological data and the nature of epidemics [2],the adopted methods was insufficient and therefore inefficient, so a new approach is used: the development of an interface between expert and computer scientist who is no longer required to start from the "whole" to achieve the "perfect" model. This interface converts the model created by the developer, to understandable language by the expert, so that he can check the validity of the model by: settings, rules, constraints, etc, and finally allow the developer to review and optimize the model, so that, any implementation of prevention and control are carried out for appropriate treatment.

The rest of the paper is organized as follows. Section 2 presents a brief review of epidemiological modeling and why do we need to translate the simulated model to the narrative languages?. A description of our model in SBML (Bio-PEPA), and how to perform its translation into narrative language, in Section 3. Section 4 describes the details of information on testing and evaluation. Section 5 summarizes the work done and also some suggestions to improve the model.

## 2. FROM NARRATIVE LANGUAGE TO A MODEL

Develop and use a good epidemiological model, remains to this day a very attractive idea, and to achieve it many researchers are struggling between having to choose the best tools and methods or to conduct a thorough training in the field in question and often they find themselves stagger between them. However, others, give little importance to neither one nor the other, rather they prefer to save their energy and adopt a technique completely original that is to transform the context expressed by an expert directly in a simulated model, as it was presented by Georgoulas and Guerriero in 2012 [3], for translating the narrative language in a "Bio-PEPA" formel model. In 2007 and 2009, Guerriero et all [4, 5] studied the translation of narrative language in a "model Beta-binders" and " a bio-inspired process calculus", the authors have assumed that it would be better to simplify the communication between experts and developers by providing a simple interface that would allow both the expert to insert their information and the developer to manipulate only the code without worrying too much about understanding everything. This approach has been baptized the passage from narrative language to a model. Although this work is regarded as a large opening in the field of modeling, however, the question arises, what happens to existing models?.

The following work, were largely inspired from Guerriero's work [5], where we suggest to do the reverse of him, kept the existing model and improve it, which means, translating the Bio-PEPA model to a narrative language.

## 3. FROM MODEL TO A NARRATIVE LANGUAGE

In order to respond to the issues raised in the previous section and based on the principle defined above, we propose an approach whose aim is to preserve existing models and also to optimize it by allowing an incremental model implementation.

An extensive literature search, which focused on methods of modeling with both, analytical and decision support tools, as well as the translation of the model in other specific formats, for approaching the narrative language, we were able to highlight Bio-PEPA [2, 6], which is formal language based on the process algebras recommended for biochemical systems and which was perfectly suited to epidemiological field. Beyond this definition, Bio-PEPA is equipped with an extension that allows translating any model in Bio-PEPA in XML format better known as SBML (Systems Biology Markup Language).

### 3.1. Bio-PEPA (Biological Performance Evaluation of Process Algebras)

Bio-PEPA is a tool, method and language based on process algebra. These are described by mathematical formalisms used in the analysis of concurrent systems [2, 7, 8], which consist of a set of processes running in parallel, can be independent or share common tasks. As it was defined in [6], the Bio-PEPA language is 7-tuple (V, N, K, FR, Comp, P, Event) Where:

- V is a set of locations,

- N is a set of auxiliary information,

- K is a set of parameters,

• E is a set of functional rates

• Comp is the set of species

• P is the component model.

• Event is the set of events.

#### 3.1.1. Characteristics of Bio-PEPA

The main features which are provided in Bio-PEPA are:

• provides a formal abstraction of biochemical systems and further epidemiological systems.

• Allows expressing any kind of interaction law expressed using functional rates.

• Allows expressing the evolution of species and their interaction.

• Defined syntax and structural semantics based on a formal representation.

• Provides the ability to perform different types of analysis from the model (continuous time Markov chain, the stochastic simulation algorithms, differential equations).

#### 3.1.2. Bio-PEPA Syntax

As defined by [9, 6], Bio-PEPA syntax is described by:

S:= (α,k) op S:=S ; S:=S+S; S:=C where

op = $\downarrow | \ominus | \oplus | \uparrow | \odot$ And

S::= S S | S(x)

Where, S: describe the species (different types of individuals); P: the model describing the system and interaction between species. The term (α, k) op S, express that the action α is described by k rate and performed by the species S, "op" define the role of S. Op={ $\downarrow$: reactant, $\uparrow$: product, $\oplus$ : activator, $\ominus$ : inhibitor, $\odot$ : generic modifier}.

### 3.2. Systems Biology Markup Language (SBML)

SBML (The Systems Biology Markup Language) is a markup language based on XML (the eXtensible Markup Language). In essence, an XML document is divided into hierarchically structured elements from a root element. Syntactically, the elements of an XML document are marked in the document itself by opening and closing pairs of tags, each element consists of a name that specifies its type, attributes, and content (elements or text).

SBML is a set of constructions' elements specific to the systems biology, defined in an XML schema. It has been adapted to the epidemiological models.

The SBML language is divided into hierarchically structured elements which are a syntax tree of language as an XML schema.

As it was defined in section 3.1, an epidemiological model is defined in Bio-PEPA by a set of compartments, species and reactions described by rates and parameters. SBML do the same by using tags and attributes [10, 14]. Figure 1 shows the general organization of SBML's TAG which are described in the following [11]:

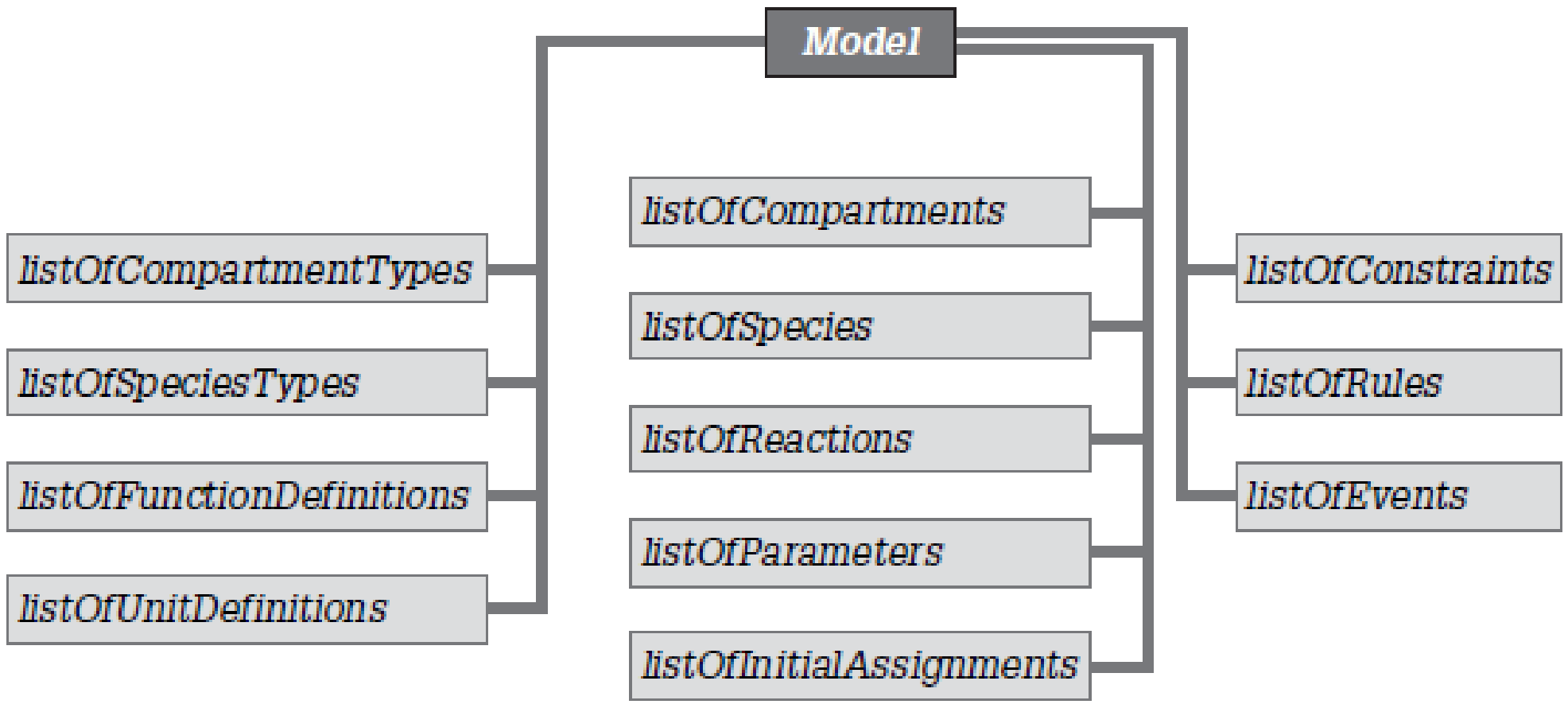

Figure 1. General organization of SBML language.

- Model : An SBML model definition consists of lists of SBML components located inside the tags : <model id="My_Model" > ….</model>.
- listOfFunctionDefinitions: The mathematical functions that can be used in the other part of the model are defined in this section.
- listOfUnitDefinitions : these units are used to explicitly specify: constants, initial conditions, the symbols in the formulas and the results of formulas.
- listOfCompartments : Is an enclosed space in which the species (species) are located.
- listOfSpecies : To specify the different entities in the model regardless of their nature, where one type of species "listOfSpeciesTypes" can be specified.
- listOfReactions : Any process whereby the transfer of a species from one compartment to another.

We specify that Representation and semantics of mathematical expressions are defined in the SBML using MathML.

### 3.3. Relation of Bio-PEPA to SBML

T The principal notions which rely Bio-PEPA to SBML are summarized in table 1, (this table was directly extracted from [2]).

Table 1: Summary of mapping from SBML to Bio-PEPA (taken from [2])

| SBML Element | Corresponding Bio-PEPA component |
|---|---|
| List of Compartments | Bio-PEPA compartments |
| List of Species | Species definitions (Name, initial concentration and compartment). Step size and level default to 1. Also used in species sequential component definitions. |
| List of Parameters | Bio-PEPA parameter list. Local parameters renamed to include re- action name. |
| List of Reactions | Species component definitions and model component definition. |
| Kinetic-Laws | Bio-PEPA Functional rates |

## 4. IMPLEMENTATION

For implementing our approach, we resumed work that we had already started in [6] which was to reproduce the spread and vaccination protocol of chickenpox in Bio-PEPA, as shown in Figure 2.

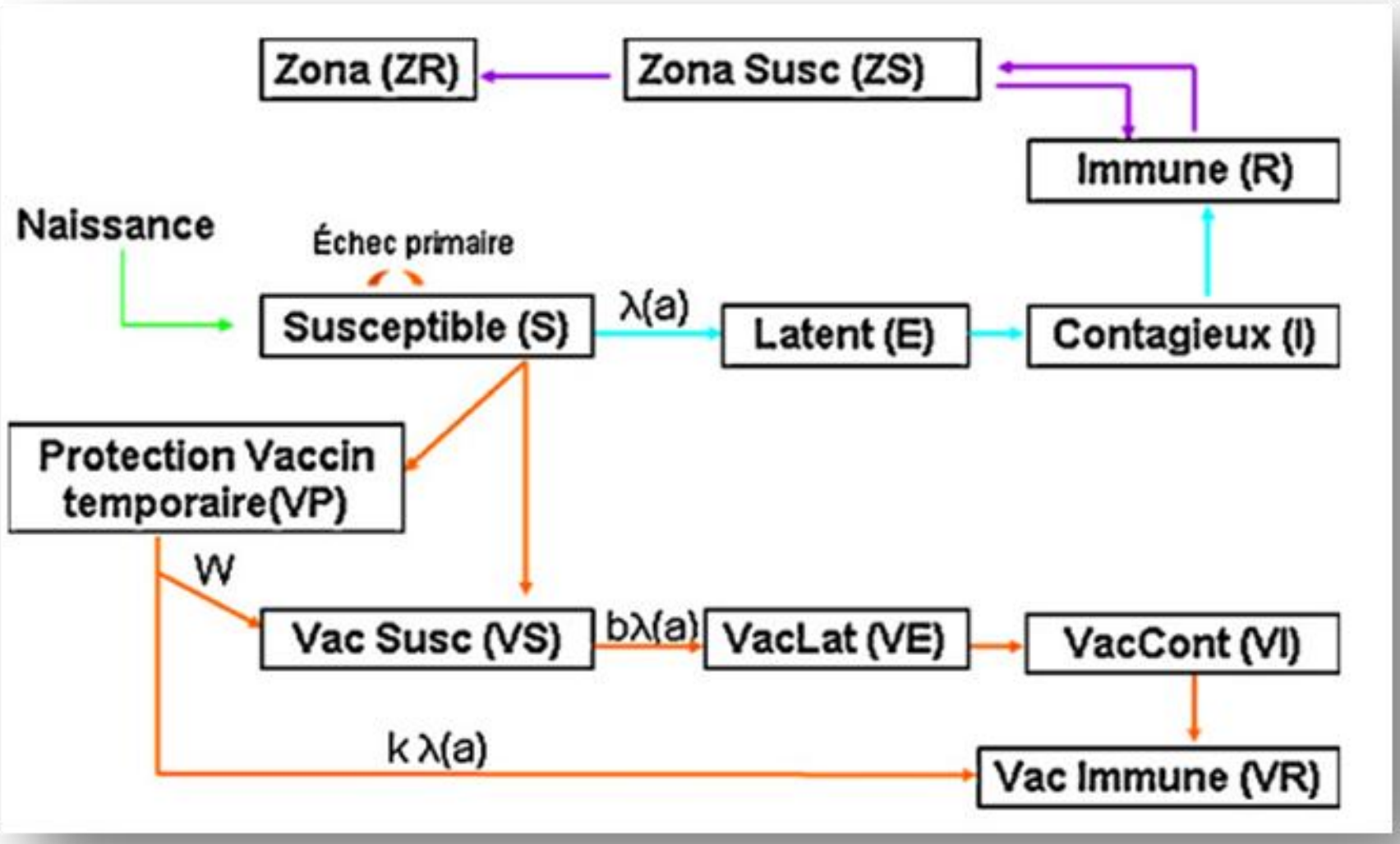

Figure 2. Model structure (taken from [12]).

The overall scheme of our approach is defined by three main steps:

- Formulation of the epidemic model in Bio-PEPA: definition of species and reactions.
- Exporting SBML file.

• Representation in narrative language: analysis of SBML file, displaying a detailed report, validation by the expert.

## 4.1. Description of model structure

Our approach, as it was structured, allows us to share our work in two main stages, the first is to develop a model with Bio-PEPA (Formulation from epidemic model in Bio-PEPA), a work that has already been done [6] and demonstrated the importance of using such a tool.

The second part (Exporting SBML file from Bio-PEPA, Representation of SBML text in narrative language) is developing a module that would translate the Bio-PEPA code in a language understood by the expert, who may easily check whether the contents of the model is adequate to the example and thus validate it.

### 4.1.1. Chickenpox model in Bio-PEPA

To better understand the modeling process, we have taken and explained in this section the most important part of the code Bio-PEPA model of chickenpox [12]. (For clarity of the document, we have listed a few parts of model).

1. **Location:** To explain the seven age groups of the model, we have represented it as compartments.

```
location Age1 in world : size = sizeAge, type = compartment;
.........
location Age7 in world : size = sizeAge, type = compartment;
```

2. **Fonctionnel rate**: Describes the interaction law between compartments.

**Exposition** = λ . S . I; describes the contact between susceptible (S) and infected (I) with λ rate.

…….

**LostVaccin** = W . VP; defines the rate of immune lost (W) of those protected by vaccination (VP).

3. **The species :** are the system entities expressed by operations describing their evolution.

**S = [(Exposition,1)↓ S + (Vaccination_1,1) ↓ S + (Vaccination_2,1)↓S;** explains what happens if executes Exposition function, Vaccination_1 or Vaccination_2.

Some lines of Bio-PEPA code are shown in Figure 3. We could note that even if the Bio-PEPA language is simple and easy for developer, however, remains an ambiguous part face which is set epidemiologist who cannot verify the validity of information represented by the developer, as the epidemiologist cannot understand the Bio-PEPA code.

We can extract from this figure, two important points, firstly the representation of chickenpox model in Bio-PEPA (right of the figure), and another hand the results of a simulation graph summarizing the status of various species (left of the figure).

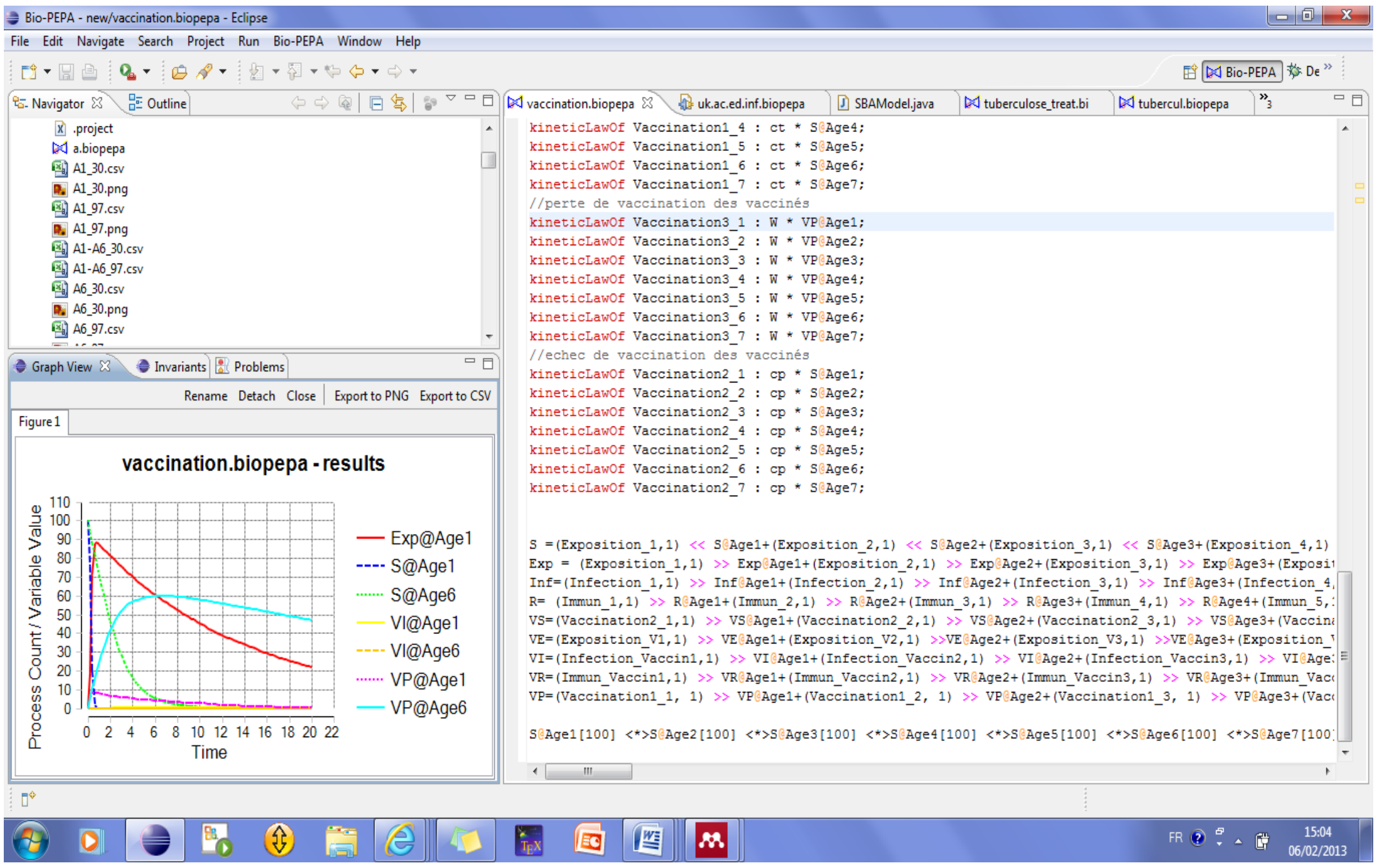

Figure 3. Global view of Bio-PEPA model.

### 4.1.2. Exporting SBML file from Bio-PEPA

Bio-PEPA provides the ability to export the model as an SBML file. As shown in Figure 4: The resulting text describes all the tags and attributes as they were presented in Section 3.2, corresponding to our model of chickenpox. It should be remembered that to study an epidemic, we must take into consideration: the environment "space", time, and various other functions. SBML can express perfectly each section describing the elements defined in Bio-PEPA.

### 4.1.3. The Chickenpox model in narrative language

To work with SBML, we need to perform a literature research, about tools analyzing and interpreting this type of descriptor, the latter revealed the JDOM [13] tool.

The main features of DOM are:

- The DOM model (unlike on this point to another famous API: SAX) is a specification that has its origins in the w3C consortium.
- The DOM model is not only a multi-platform specification, but also multi-languages: as Java, JavaScript, etc.
- DOM presents documents as a hierarchy of objects, from which, more specialized interfaces are themselves implemented: Document, Element, Attribute, Text, etc. With this model, we can treat all DOM components either by their generic type, “Node”, or by their specific type “element, attribute”, many methods of navigation allow navigation in the tree without having to worry about the specific type of component Treated.

```
<?xml version="1.0" encoding="UTF-8"?>
<sbml version="3" level="2"
xmlns="http://www.sbml.org/sbml/level2/version3">
<model id= "vaccination_biopepa">
<listOfCompartmentTypes> <compartmentType id="Compartment"/>
<compartmentTypeid="Membrane"/>
</listOfCompartmentTypes>
<listOfCompartments>
<compartmentid="Age7" outside="world" size="100000.0"
compartmentType="Compartment"/>
……..
<compartment id="Age5" outside="world" size="100000.0"
compartmentType="Compartment"/>
</listOfCompartments>
<listOfSpecies> <species id="Exp_Age1" hasOnlySubstanceUnits="true"
substanceUnits="item" compartment="Age1" name="Exp"/>
……
<species id="VS_Age7" hasOnlySubstanceUnits="true"
substanceUnits="item" compartment="Age7" name="VS"/>
</listOfSpecies>
```

Figure 4. Chickenpox model in SBML.

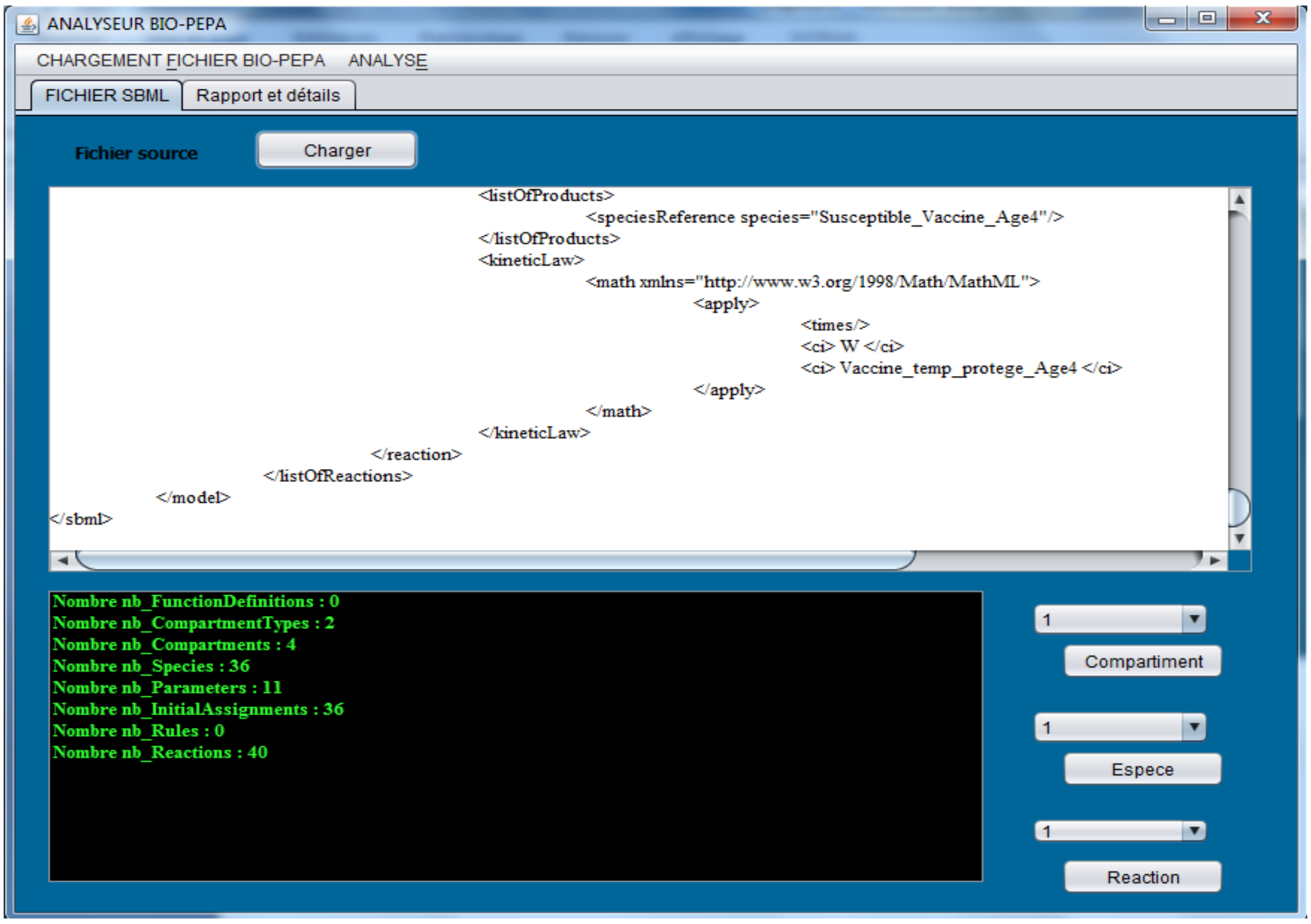

Figure 5. (a) From SBML code to narrative language.

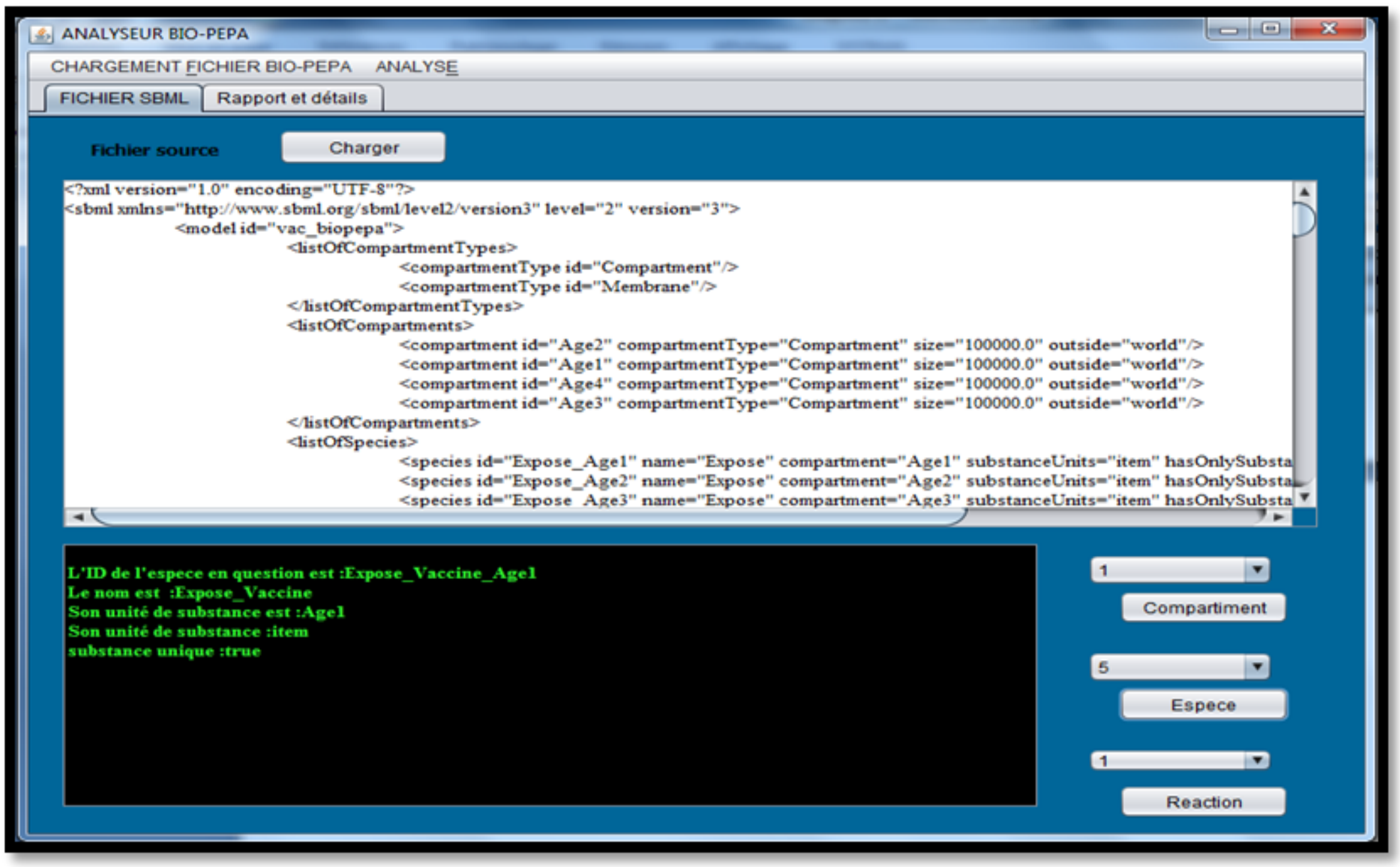

Figure 5 (b) From SBML code to narrative language.

Figures 5 (a, b) shows the interface of our application based on JDOM model and thus gathering the steps defined above. White space viewed in the figure corresponds to the loading of SBML file, when the black area corresponds to the translation and analysis of SBML in narrative language understandable by the expert, the expert in this way has no difficulty in verifying the validity of the model. The user-friendly interface allows him to navigate the various components of the model (species, function of interaction, locations ... etc.)

To validate our application, we made a change in the initial code (Bio-PEPA) where we intentionally caused an error in our model, the generation of the latter, as shown clearly in Figure 6 that the expert could detect the species and reactions which are missing, and therefore it can easily report them to us. (The red frame line specifies the error caused by the number of species missed).

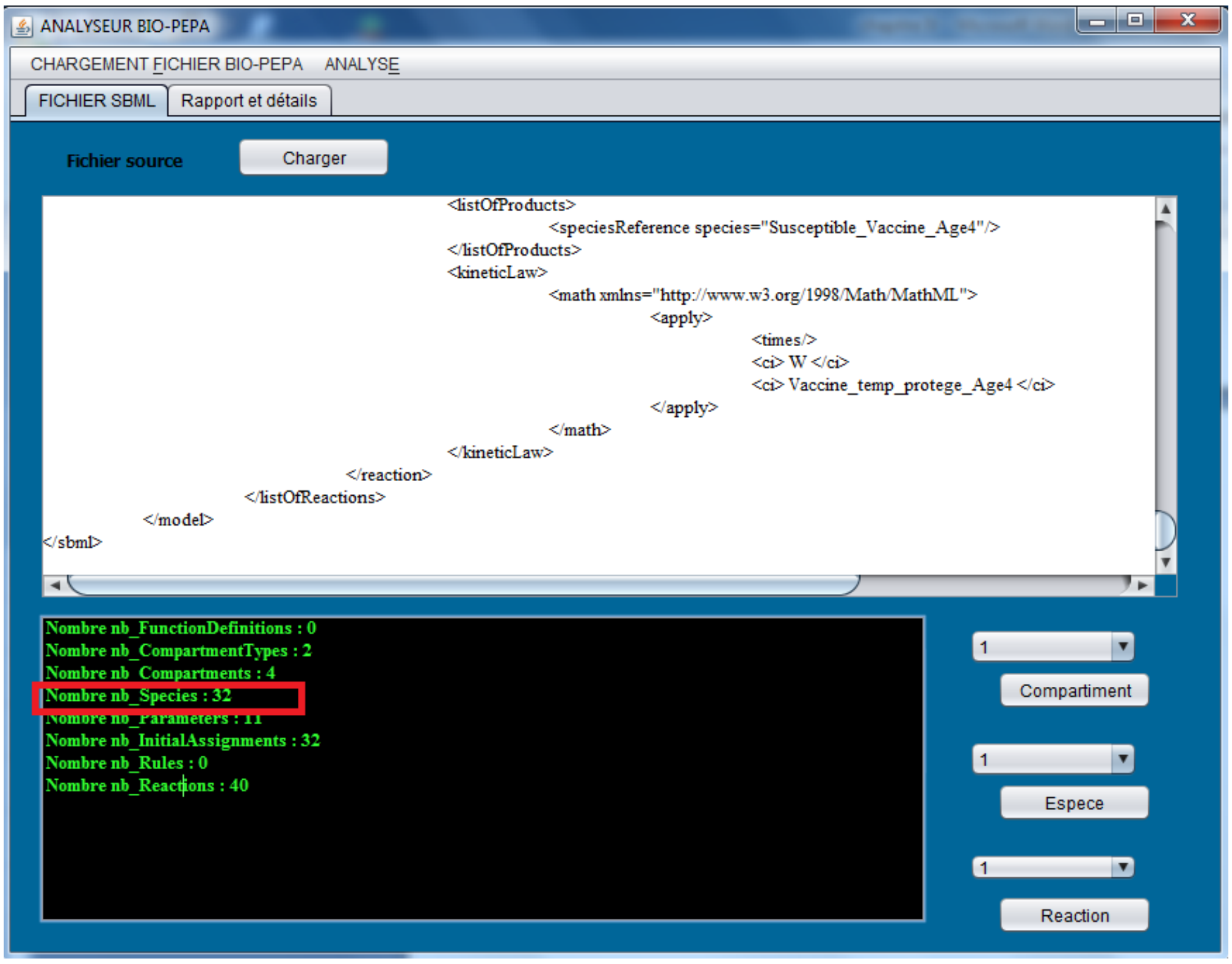

Figure 6. Detection error after translatation the model into narrative language.

## 5. Conclusions

Modeling and simulation are very useful to understand and predict the dynamics of various biological phenomena. The Bio-PEPA approach seems to be an interesting and powerful approach to address such problems. Through its various features it allows an easy development of the computer model and a transparent way for biologists between the real system and the built which helps a faithful representation of the phenomenon studied. Nevertheless, in case of occurrence of a new event, which has been badly treated by the developer and therefore omitted, correction model is considered a tedious task for both. This is the reason for which, we have introduced a new module (interface), where the expert can easily detect this omission and thus back to the developer, who may discern error and quickly position it, on the Bio-PEPA model.

As perspective to strengthen this work, why not attached it to the one that was mentioned in Section 2, and thus drift toward a cyclical pattern, which would not require even the presence of developer, however, after reflection, what will become the expert, front of its multitude of information? After a brief literature review the idea of integrating it with the world of data mining would be a much better idea to fruition.